\pdfoutput = 1
\documentclass[aps,pra,superscriptaddress,twocolumn,reprint,longbibliography,floatfix]{revtex4-1}
\usepackage{amsmath}
\usepackage{amsfonts}
\usepackage{graphicx}
\usepackage{xcolor,color}
\usepackage{amssymb,graphicx}
\usepackage{amsthm}
\usepackage{enumerate}
\usepackage{float}
\usepackage{tikz}
\usepackage{pgfplots}
\usepackage{amsmath}
\usepackage{color}
\usepackage[colorlinks,urlcolor=blue,citecolor=blue,linkcolor=blue,pdfstartview=FitH]{hyperref}

\usetikzlibrary{external}
\pgfplotsset{compat=1.3,scaled y ticks=false}
\usetikzlibrary{calc}

\definecolor{col_omega_1_over_200}{RGB}{255, 0, 0}
\definecolor{col_omega_1_over_33}{RGB}{140,140,140}
\definecolor{col_omega_1_over_10}{RGB}{0, 0, 255}
\definecolor{col_omega_1_over_6}{RGB}{247, 189, 0}
\definecolor{col_omega_1_over_3}{RGB}{158, 125, 16}
\definecolor{col_omega_1}{RGB}{0, 255, 0}
\definecolor{col_omega_10}{RGB}{255, 171, 0}
\definecolor{col_omega_100}{RGB}{255, 173, 250}

\definecolor{myred}{RGB}{140, 0, 0}
\definecolor{myorange}{RGB}{255, 157, 0}
\definecolor{myyellow}{RGB}{246, 255, 0}
\definecolor{mylime}{RGB}{162, 255, 0}
\definecolor{mygreen}{RGB}{17, 153, 0}
\definecolor{mylightblue}{RGB}{4, 212, 198}
\definecolor{mydarkblue}{RGB}{14, 0, 54}
\definecolor{mydarkcyan}{RGB}{1, 123, 125}
\definecolor{mydarkgreen}{RGB}{20, 128, 1}
\definecolor{mydarkorange}{RGB}{138, 92, 0}
\definecolor{mypurple}{RGB}{138, 0, 133}
\definecolor{mypink}{RGB}{255, 173, 250}
\definecolor{palered}{RGB}{255, 191, 163}
\definecolor{paleblue}{RGB}{227, 252, 251}
\definecolor{sunorange}{RGB}{235, 188, 101}
\definecolor{magenta}{RGB}{255, 0, 255}
\definecolor{grey}{RGB}{140,140,140}
\definecolor{mydarkbrown}{RGB}{81,51,0}
\definecolor{brown}{RGB}{255, 174, 0}
\definecolor{mustard}{RGB}{255, 203, 0}
\definecolor{ginger}{RGB}{179, 111, 12}
\definecolor{palered}{RGB}{255, 191, 163}
\definecolor{sunorange}{RGB}{235, 188, 101}
\definecolor{grey}{RGB}{140,140,140}
\definecolor{mustard}{RGB}{255, 203, 0}

\begin{document}
\title{Periodic quenches across the Berezinskii-Kosterlitz-Thouless phase transition}
\author{K. Brown}
\affiliation{Joint Quantum Centre Durham-Newcastle, School of Mathematics, Statistics and Physics, Newcastle University, Newcastle upon Tyne, NE1 7RU, United Kingdom}
\author{T. Bland}
\affiliation{Joint Quantum Centre Durham-Newcastle, School of Mathematics, Statistics and Physics, Newcastle University, Newcastle upon Tyne, NE1 7RU, United Kingdom}
\author{P. Comaron}
\affiliation{Joint Quantum Centre Durham-Newcastle, School of Mathematics, Statistics and Physics, Newcastle University, Newcastle upon Tyne, NE1 7RU, United Kingdom}
\affiliation{Institute of Physics Polish Academy of Sciences, Al.~Lotnik\'ow 32/46, 02-668 Warsaw, Poland}
\author{N. P. Proukakis}
\affiliation{Joint Quantum Centre Durham-Newcastle, School of Mathematics, Statistics and Physics, Newcastle University, Newcastle upon Tyne, NE1 7RU, United Kingdom}

\date{\today}

\begin{abstract}
The quenched dynamics of an ultracold homogeneous atomic two-dimensional Bose gas subjected to periodic quenches across the Berezinskii-Kosterlitz-Thouless (BKT) phase transition are discussed. Specifically, we address the effect of periodic cycling of the effective atomic interaction strength between a thermal disordered state above, and a highly ordered state below the critical BKT interaction strength,
by means of numerical simulations of the stochastic projected Gross-Pitaevskii equation. 
Probing the emerging dynamics as a function of the frequency of sinusoidal driving  from low to high frequencies reveals diverse dynamical features, including phase-lagged quasi-adiabatic reversible condensate formation, resonant excitation consistent with an intrinsic system relaxation timescale, and gradual establishment of dynamically-recurring or time-averaged non-equilibrium states with enhanced coherence which are neither condensed, nor thermal.
Our study paves the way for experimental observation of such driven non-equilibrium ultracold superfluid states.
\end{abstract}

\maketitle

\section{Introduction}\label{sec:intro}

The quench dynamics of a quantum system across a phase transition are an exciting subject of active ongoing research~\cite{bray_theory_1994,dziarmaga_review_2010,polkovnikov_review_2011,del_campo_universality_2014,proukakis_snoke_littlewood_2017}. Controlled studies have been performed in a plethora of diverse systems, including spin systems, superconductors, superfluids, ultracold atoms and exciton-polariton systems~\cite{KZnum-a,KZnum-b,KZnum-c,KZnum-d,KZexp-a,KZexp-b,KZexp-c,KZexp-d,KZexp-e,KZexp-f,KZexp-g,KZexp-gg,weiler_08,erne_universal_2018}.
Most studies to date have focussed on a simplified scenario, whereby the system is driven once across the phase transition by a time-dependent external control parameter such as chemical potential, temperature, or pumping laser \cite{weiler_08,zurek_09,jelic2011quench,damski_10,das_12,liu_18,zamora2020kibble,comaron_19,ComaronPRLpolaritons}.
The particular case of linear quenching is known to lead to the established Kibble-Zurek scaling law, which quantifies the dependence of correlation functions and spontaneous emergence of defects on the quench rate. Such a model, first proposed in the cosmological context \cite{kibble1976topology} and subsequently carried through to the condensed matter realm \cite{zurek1985cosmological}, has since been extensively studied numerically~\cite{del_campo_universality_2014,KZnum-a,KZnum-b,KZnum-c,KZnum-d,weiler_08,damski_10,das_12,Su13,Matuszewski2014,McDonald2015,liu_18,liu_20,bland_marolleau_20,liu_20,zamora2020kibble}, and in a broad range of experimental settings~\cite{KZexp-a,KZexp-b,KZexp-c,KZexp-d,KZexp-e,KZexp-f,KZexp-g}, with experimental studies in ultracold atoms addressing the effect across different dimensionalities and geometries 
\cite{KZexp-gg,weiler_08,das_12,lamporesi2013spontaneous,corman2014quench,navon2015critical,chomaz2015emergence,braun2015emergence,donadello2016creation,liu_18,ko2019kibble,keesling2019quantum}.

The related question of how the presence of a sinusoidally-modulated driving across a phase transition may affect the system dynamics arises naturally.
Such a periodic cycling was experimentally investigated in the context of three-dimensional (3D) harmonically trapped ultracold atoms, where a time-dependent dimple microtrap was used to controllably induce a periodic phase space modulation, leading to a reversible cycling across the thermal and the Bose-Einstein condensation phase \cite{stamperkurn1998reversible}.
Condensate formation was found to lag behind the applied constant-frequency
modulation of the laser power. Such findings were subsequently reproduced
 qualitatively by means of numerical simulations based on the stochastic Gross-Pitaevskii equation~\cite{stoof2001dynamics}.
Motivated by the above pioneering works~\cite{stamperkurn1998reversible,stoof2001dynamics}, and by our recent studies of instantaneously quenched two-dimensional quantum gases~\cite{comaron_19,groszek2020crossover}, here we address the corresponding periodically driven phase transition crossing in the context of (quasi) two-dimensional homogeneous ultracold atomic Bose gases. Two-dimensional systems are interesting in their own right, due to the different nature of the underlying Berezinskii-Kosterlitz-Thouless (BKT) phase transition~\cite{berezinskii1972destruction,kosterlitz1973ordering}, associated with binding-unbinding of vortex-antivortex pairs, previously observed in diverse physical contexts~\cite{resnick1981kosterlitz,PhysRevLett.40.1727,safonov1998observation,nitsche2014algebraic}.
Such properties have also been studied in ultracold atomic systems both experimentally ~\cite{stock2005,hadzibabic2006berezinskii,kruger2007p,clade2009observation,PhysRevLett.105.230408,hung2011observation,yefsah_11,LiChung2013,Fletcher2015,chomaz2015emergence}, and theoretically~\cite{prokofev2001critical,Prokofev2002,davis_01,simula_06,Giorgetti2007,Holzmann2008,Simula2008,Bisset2009,foster2010vortex,holzmann_10,mathey2010light,Cockburn_2012,mathey_17,Karl2017,gawryluk_19,comaron_19} with the transversal (typically harmonic) confinement in the other direction offering a way to control the effective two-dimensional interaction experienced between the atoms.

In this work, we perform a detailed quantitative analysis of the role of driving frequency on the cyclic phase transition crossing, focussing on the particular case of a periodically-driven 2D homogeneous ultracold atomic Bose gas.
This is facilitated by a periodically-modulated-in-time interaction strength, between an initial incoherent state close to, but above, the BKT phase transition, and a state with lower interaction strength below the BKT phase transition exhibiting a high degree of coherence. 
Our numerical study, performed at fixed system temperature, focusses in parallel on the effect of external driving on density, vortex number, momentum spectrum and coherence.
Our parameter choice is based on accessible experimental regimes, building on our earlier work based on an {\em instantaneous} interaction quench from above to below the BKT threshold, which focussed on equilibrium properties and late-time phase-ordering dynamics~\cite{comaron_19}.

Deep in the superfluid regime, where the system is highly condensed, modulations of the interaction strength have been studied both experimentally and theoretically, in different contexts.
Periodic modulations of the interaction strength between two values in the superfluid regime --  conducted within the context of the Gross-Pitaevskii equation, for which the system is assumed to be coherent, and often termed `Feshbach Resonance Management'~\cite{Kevrekidis_FRM_03} -- have demonstrated the emergence of Faraday patterns~\cite{staliunas_02} and have been used to study aspects of condensate stability and soliton dynamics~\cite{saito_FRM_03,abdullaev_FRM_03}.
Condensate experiments based on a sudden interaction modulation found  an interesting analogy with Sakharov oscillations in the early Universe~\cite{Hung_2013}, while periodic interaction strength modulations led to matter-wave jet emission (Bose fireworks)~\cite{chin_prl_18,chin_nature_17} and other interesting patterns~\cite{chin_natphys_20}. 

The novel feature of our present study is that our periodic interaction quenches are neither conducted between two states deep in the superfluid regime, nor do they start from a well-formed condensate -- but instead the interaction strength is modulated across the phase transition multiple times, in direct analogy to the work of \cite{stamperkurn1998reversible}.
For this reason, our analysis is based on the stochastic (projected) Gross-Pitaevskii equation~\cite{Bradley2008,Blakie2008,Proukakis2008,Proukakis13}.

Beyond the expected regimes of extremely slow pumping (which allows the system to proceed adiabatically through instantaneous equilibrium states) and  extremely rapid pumping which only mildly perturbs the initial state, we 
find the periodic driving to be intrinsically resonant with a characteristic relaxation time, at which the periodic modulation of the scattering length causes strongly-non-equilibrium features to emerge. This resonance separates two interesting distinct and experimentally relevant physical regimes: Driving frequencies lower than the resonant value lead to reproducible dynamics which are independent of the quench cycle, and -- while not fully equilibrated -- resemble certain features of the corresponding-parameter equilibrium states.
This regime bears close analogy to the experimental 3D findings of Ref.~\cite{stamperkurn1998reversible} (upon excluding observed atom losses). In the opposite regime of frequencies exceeding the resonant value, the system grows gradually over multiple quench cycles, and can -- for frequencies few times the resonant frequency -- accommodate significant coherence.

This paper is structured as follows:
After presenting a brief overview of the 3D experiment of Ref.~\cite{stamperkurn1998reversible}, we outline the numerical scheme and quench protocol to be used in our work (Sec.~II). We start by discussing the dependence of the emerging density and vortex dynamics on driving frequency, and present a detailed quantitative analysis of their respective dynamical time-delay with respect to the external periodic driving (Sec.~III): this allows us to identify an intrinsic system frequency, and thus to focus our subsequent analysis  on the physically interesting and experimentally relevant regime of driving frequencies which are up to 20 times slower, or faster, than the identified resonant frequency. Within this experimentally interesting range where most novel dynamical features emerge, we analyse the dependence of the momentum spectrum on the driving frequency and identify parameter regimes of high coherence in spite of fast driving (Sec.~IV). We then study the maximum attained coherence as a function of quench frequency, and the emerging non-equilibrium steady-state under rapid pumping (Sec.~V), before concluding.

\section{Quench Protocol and Modelling Scheme}\label{sec:quench}




\subsection{Experiment of Stamper-Kurn {\em et al.} \cite{stamperkurn1998reversible}}

In a ground-breaking paper, Stamper-Kurn \textit{et al.}~\cite{stamperkurn1998reversible} achieved reversible condensate formation in a three-dimensional gas of sodium atoms. Initially, this gas was confined to a broad harmonic trap and cooled to just above the critical temperature, $T_c$, at which condensation onsets. Subsequently, a thin well (dimple trap) was introduced at the centre of the trap, using an infrared laser. This granted the sodium atoms access to a new, lower-energy state. The well was made so steep that it could only contain a single new energy level. As well depth was increased, by increasing the strength of the laser, the energy of this single state decreased to the extent that a condensate formed in this well. The condensate fraction grew with well depth, and hence with laser power. The power of the laser beam was then sinusoidally modulated between 0 and 7mW at a frequency of 1Hz, which resulted in the periodic cyclic growth and decay of the condensate fraction, $N_0/N$, from 0 to 6$\%$.
A non-zero condensate fraction was observed to recur for 15 oscillations, even though the peak condensate fraction decayed slightly with each oscillation.
This decay was shown to arise from atom loss, rather than as a consequence of this periodic crossing of the phase transition. The latter was confirmed by the fact that the same decay in the peak values was observed even when the power of the laser beam was held constant. Importantly, a $\sim70$ms time delay was observed between the times of maximum laser power and corresponding time of peak condensate fraction, giving some insight into the condensate formation time under this protocol.

To interpret the observed findings, Stoof and Bijlsma \cite{stoof2001dynamics} pioneered the use of a stochastic Gross-Pitaevskii equation as a model to study reversible condensate formation. Introducing a steep well in the potential, as done experimentally, is equivalent to altering the effective chemical potential, $\mu_{\textrm{eff}}$, of the system. Hence, in this numerical investigation, the reversible formation of a one-dimensional condensate was simulated by modulating the value of $\mu_{\textrm{eff}}$ as $\mu_{\textrm{eff}}(t) = \mu\textrm{sin}(\omega_D t)$, where $\mu$ is the chemical potential of the system in the absence of a well and $\omega_D$ is the frequency of the imposed periodic modulation in $\mu_{\textrm{eff}}(t)$. Stoof and Bijlsma analysed the evolution of the condensate density at the trap centre and qualitatively reproduced the delay in condensate growth observed by Stamper-Kurn \textit{et. al}~\cite{stamperkurn1998reversible} (but observing a slightly longer  time of $\sim$100ms, which could be attributed to the different probed dimensionality and potential uncertainties of exact experimental parameters). That study provided the first numerical evidence of a repeated phase transition crossing in ultracold atomic gases. Since that work, the stochastic Gross-Pitaevskii model introduced by Stoof \cite{stoof2001dynamics}, and its closely-related stochastic projected Gross-Pitaevskii equation \cite{Bradley2008,Blakie2008}, have been used to study a broad range of dynamical ultracold phenomena~\cite{Blakie2008,Proukakis2008,Proukakis13,berloff_brachet_14}, including the study of quenched phase transitions
across different geometries,  dimensionalities and mixtures~\cite{Proukakis03,Proukakis_Schmiedmayer_2006,weiler_08,Proukakis09,zurek_09,damski_10,das_12,Su13,De_2014,McDonald2015,Liu_2016,gallucci2016engineering, Kobayashi_16a,Kobayashi_16b,Eckel_2018,liu_18,Ota_2018,comaron_19,bland_marolleau_20,liu_20}, with direct successful modelling of phase transition experiments~\cite{weiler_08,liu_18,bland_marolleau_20}.

\subsection{Numerical Model and Parameter Regime}


In this work we analyse the quenched system dynamics in terms of the 
Stochastic Projected Gross-Pitaevskii Equation (SPGPE), which describes the `classical' field, $\Phi({\mathbf r},t)$, of all highly-populated modes up to a fixed energy cut-off. Its dynamics are governed by~\cite{Blakie2008}:
\begin{align}
    i\hbar\frac{\partial\Phi}{\partial t} = \hat{\mathcal{P}}\Big\{ \Big(1- i\gamma\Big)\Big[-\frac{\hbar^2}{2m} \nabla^2  +  g|\Phi|^2 - \mu \Big]\Phi + \eta\Big\}\,.
    \label{eqn:spgpe}
\end{align}
Here $-i \gamma$ corresponds to a dissipative/growth term, arising from the coupling of the `classical field' modes $\Phi$ to the high-lying modes, which are treated as a heat bath. Consistent with other treatments~\cite{comaron_19,Ota_2018}, we set  $\gamma = 0.01$ in our current calculations, 
 noting that typical physical evolution timescales are set by the scaled time $\gamma t$, controlling system dynamics and growth.
The long-term evolution/steady-state is determined by the balancing of the kinetic energy contribution and the nonlinear interaction term $g |\Phi({\mathbf r}, t)|^2$ with the bath chemical potential $\mu$. 
The presence of the dynamical noise term $\eta$, associated with the collisional randomness of growth/decay processes differentiates each numerical run in a manner analogous to experimental shot-to-shot fluctuations. The noise correlations are  given by $\langle \eta^{*}(\textbf{r},t)\eta(\textbf{r}',t') \rangle = 2\hbar\gamma k_{\textrm{B}}T\delta(t-t')\delta(\textbf{r}-\textbf{r}')$.
Finally, $\hat{\mathcal{P}}$ is a projector which constrains the dynamics of the system within a finite number of macroscopically occupied modes, up to an ultraviolet energy cutoff 
	$\epsilon_\mathrm{cut} (\mu,T) =k_\mathrm{B}T\log(2)+\mu$,
with the mean occupation of the last included mode set to be $\sim 1$. This relation sets our numerical grid spacing to $\Delta x \le \pi/\sqrt{8m\epsilon_\text{cut}}$ \cite{Blakie2008}.

In this work, we consider a two-dimensional, weakly interacting, homogeneous Bose gas, confined to a square box of sides $L_x = L_y = 100 \mu\textrm{m}$. 
Considering experimentally accessible regimes, our presented analysis is based on a gas of ultracold \textsuperscript{39}K bosons with mass ${m = 6.47 \times 10^{-26}}$ kg, a chemical potential $\mu = 1.92 k_B$nK and a temperature, $T = 50$nK \cite{footnote1}.
The corresponding 2D interaction strength is given in terms of the tight transversal harmonic confinement of frequency $\omega_{\perp}$ by 
${g_{2D}= g_{3D} / \sqrt{2\pi}\ell_\perp}$, where $g_{3D} = 4 \pi \hbar^2 a_{s}/m$, $a_s=7.36$nm is the background s-wave scattering length \cite{derrico2007feshbach}, and ${\ell_\perp=\sqrt{\hbar/m\omega_\perp}}$ the transverse harmonic length.
This yields a dimensionless coupling constant
$g = m g_{2D} / \hbar^2 = \sqrt{8\pi}{a_s}/{\ell_\perp}$.

The location of the BKT phase transition is fixed by the relation \cite{prokofev2001critical}
\begin{equation}
\label{eqn:g_crit}
\frac{\mu}{k_\mathrm{B}T} \bigg\rvert_\mathrm{BKT} \approx \frac{g_c}{\pi} \ln\left(\frac{C}{g_c}\right),
\end{equation}
where $g_c$ denotes the critical value of the dimensionless strength at the BKT transition, and Monte-Carlo analysis gives the constant $C\sim 13.2$ \cite{Prokofev2002}.
This relation sets the value of $g_c=1.83 \times 10^{-2}$ for our parameters.
This coupling constant can be varied by changing the transverse confinement, and the parameters considered in this work relate to the range 
$1.83 \times 10^{-3} < g < 3.48 \times 10^{-2}$. Implicit in our description is the modulation of the scattering length through Feshbach resonance \cite{chin2010feshbach}, or the transverse confining harmonic oscillator frequency $\omega_\perp$. 
All presented results correspond to an average over $\mathcal{N} = 50$ stochastic realizations, which is large enough such that the error bars on all figures are of the same size order as the marker itself.




\subsection{Quench Protocol}

Stimulated by the work of Refs.~\cite{stamperkurn1998reversible,stoof2001dynamics}, we impose here a symmetric interaction quench, as follows:
Firstly, we allow the system to dynamically equilibrate close to, and on the disordered side of, the BKT critical region.
Specifically we choose to equilibrate initially to a value $g=1.9 g_c$ (which corresponds to an initial number of $N=3\times10^4$ atoms in the classical field). We verify that this state is in equilibrium via calculation of the first order correlation function, which exhibits the correct exponential decay law, and ensuring the vortex number reaches a steady state value.
We then initiate a periodic quench in the interaction strength which is symmetric about the critical point, via
\begin{align}
     g(t) = g_c(1+\alpha\textrm{cos}(\omega_D t))\,,
     \label{eqn:gt}
\end{align}
where, $\omega_D$ is the driving frequency of the quench and $\alpha$ is the constant periodic amplitude. Based on our chosen initial state, we fix the modulation amplitude to $\alpha = 0.9$, implying that the quench proceeds between an initial state with $g=1.9 g_c$ (corresponding to $T>T_\text{BKT}$), and a final highly-ordered state with $g=0.1 g_c$ ($T<T_\text{BKT}$)~\cite{footnote2}.

Throughout this work, we examine how the system behaviour varies for different driving frequencies in the range $\omega_D = 2 \pi \times [1/40 ,\, 500]$Hz, which cover the entire behaviour from quasi-adiabatic ($ \omega_D = 2 \pi \times (1/40)$Hz) to extremely rapid ($ \omega_D = 2 \pi \times 500$Hz), which fully encompass the entire range of potential parameters for experimental investigation.

%
%
%
In presenting our results, we introduce the time-dependent deviation from criticality
\begin{align}
    \Delta g(t) = \frac{g(t)-g_c}{g_c},
    \label{eqn:delg}
\end{align}
which is constrained to oscillate between $+\alpha$ (disordered) and $-\alpha$ (ordered).

\begin{figure}
\centering
\includegraphics[width=0.9\columnwidth]{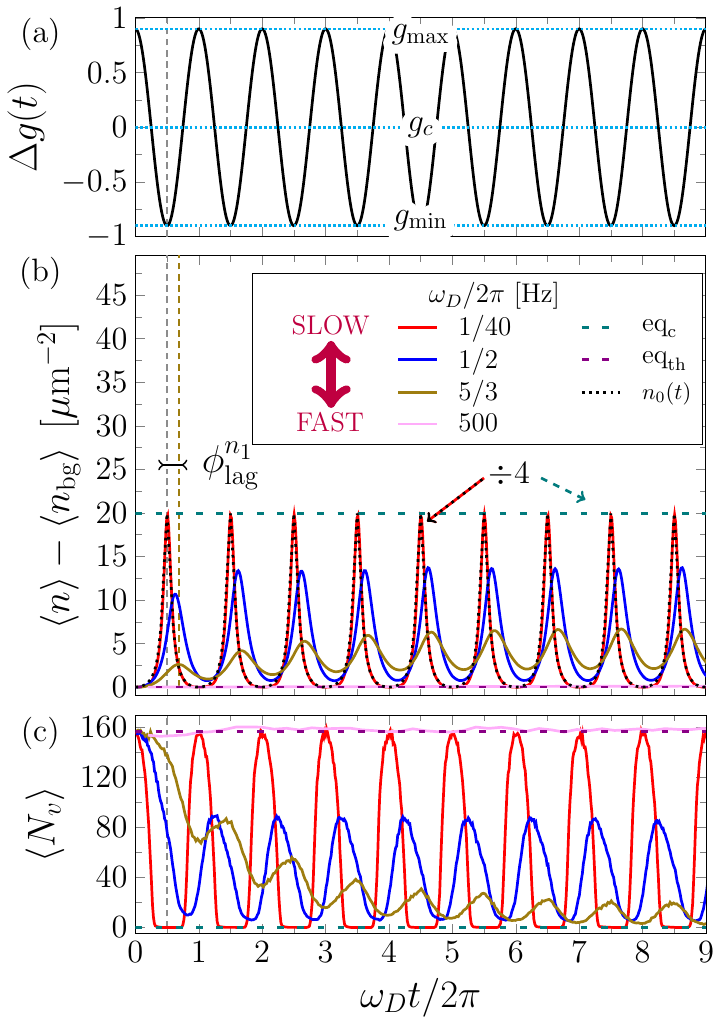}
    \caption{Repeated quenches through the BKT phase transition, with time normalised to the quench period through the time scale $2\pi/\omega_D$. (a) Quench protocol [Eq.~\eqref{eqn:delg}]. Dotted line through $\Delta g = 0$ separates the incoherent (thermal) ($\Delta g >0$) and (partly) superfluid ($\Delta g <0$) states.  (b) Evolution of the  c-field density averaged over space and stochastic realisations, compared to its initial value $\langle n(t=0) \rangle \equiv \langle n _{\textrm{bg}} \rangle$ for various driving frequencies $\omega_{D}$. Also shown are the equilibrium densities for the initial thermal state ($\text{eq}_\text{th}$, lower purple horizontal line) and the final-parameter part-superfluid system $\text{eq}_\text{c}$ (upper blue horizontal line) respectively, found by fixing $g(t)=g_\text{max}$ ($\Delta g=0.9$) and $g(t)=g_\text{min}$ ($\Delta g=-0.9$), and the equilibrium density $n_0(t)=\mu/g(t)$. The slowest quench and condensate equilibrium data are reduced by a factor of 4 for visual aid. For a given driving frequency, the phase lag between the $i_{\textrm{th}}$ trough of $\Delta g$ and the $i_{\textrm{th}}$ peak of the average density is denoted $\phi_{\textrm{lag}}^{n_i}$. The vertical grey dashed line throughout all three panels highlights the time of the first trough in $\Delta g$. 
    (c) Evolution of the average number of vortices, $\langle N_{v} \rangle$, for the same quench frequencies and equilibria as above.}
    \label{fig:quenchprot}
\end{figure}

%



This quench protocol can be easily visualized in Fig.~\ref{fig:quenchprot}(a).
Positive values of $\Delta g$ indicate that $g > g_{\textrm{c}}$, with the system in the disordered (thermal) phase, while $\Delta g < 0$ indicate that $g < g_{\textrm{c}}$, which initiates the (quasi)-condensate formation process.
For better understanding, in what follows we also briefly compare our dynamical periodic quench results to the corresponding equilibrium regimes in the two limiting cases of our quench cycle. 

\section{Dynamical Results}\label{sec:dyn}

The system dynamics under periodic quenching can be characterized in terms of its density, vortex number, occupation spectrum and coherence, to which we next turn our attention. 

\subsection{Density and Vortex Delayed Oscillations}

We start our analysis by considering the impact of the periodic interaction strength modulation shown in Fig.~\ref{fig:quenchprot}
on density and vortex number. 

Specifically, Fig.~\ref{fig:quenchprot}(b) shows the evolution of the average modulation $\langle n \rangle - \langle n _{\textrm{bg}} \rangle$ of the c-field density $\langle n \rangle$, compared to its initial value $\langle n _{\textrm{bg}} \rangle$, averaged over space and stochastic realisations, for a range of driving frequencies, while  
Fig.~\ref{fig:quenchprot}(c) shows the corresponding average evolution of the vortex number in the system.
It is well-known that above the BKT phase transition ($\Delta g>0$) the phase of the system is random, corresponding to an exponentially-decaying phase correlation function: this incoherent state can be thought of as a system with a large number of `free' vortices, whose exact number is fixed by the system size, grid resolution, atomic mass, temperature and interaction strength (the number itself is not important). 
As we scan the system across different equilibrium configurations, we find that the number of vortices at equilibrium decreases abruptly across the BKT critical region, reaching a very low, or even zero value as the system approaches its pure superfluid limit ($g \ll g_c$)~\cite{foster2010vortex,gawryluk_19,comaron_19}.

Let us first consider the limiting (idealized) cases of extremely slow and rapid driving:
For extremely slow driving, we expect the system to evolve adiabatically through corresponding equilibrium states, oscillating between the initial purely incoherent state with $\langle n \rangle = \langle n _{\textrm{bg}} \rangle$ which is filled by vortices (here $\langle N_{v} \rangle \sim 160$), and a higher value ($\langle n \rangle - \langle n _{\textrm{bg}} \rangle \approx 80\mu$m$^{-2}$ ), which corresponds to the equilibrium density value at $g=g_\text{min}=0.1g_c$, and $T=50$nK, with such state having $\langle N_{v} \rangle = 0$. The timescale for such periodic change would be set by the slow drive frequency $2 \pi / \omega_D$, in the limit $\omega_D \rightarrow 0$. In fact, the equilibrium density is simply given by the time-dependent homogeneous solution $n_0(t)=\mu/g(t)$, which the idealized case $\omega_D\to0$ follows in Fig.\ref{fig:quenchprot}(b).
The corresponding averaged densities and vortices in the two limiting cases are shown Fig.\ref{fig:quenchprot}(b)-(c) by the horizontal purple (initial incoherent state with $g=g_{\rm max}$) and blue (part-superfluid state with $g=g_{\rm min}$) lines. 

In the opposite extreme limit of very rapid driving ($\omega_D \rightarrow \infty$), the system has no time to adjust to the driving parameters, and remains in an effective incoherent steady-state close to the initial state, i.e. its density and vortex number closely mimic the flat behaviour of the purple horizontal lines, as shown here for $\omega_D/2 \pi = 500$Hz.

Neither of those regimes is intrinsically interesting to study, so our numerical analysis focuses on a broad range of intermediate driving frequencies, from the quasi-adiabatic to the dynamically-driven regimes identified below.

The slowest quench analyzed in detail in this work is shown by the red curves in Fig.~\ref{fig:quenchprot}(b)-(c), and corresponds to a drive frequency $\omega_D = 2 \pi \times (1/40)$ Hz. This has been selected such that both the density and vortex number closely resemble the adiabatic case (with $\langle N_{v} \rangle \approx 0$), but strictly speaking the frequency of the drive is such that the state reached at $g_{\rm min}$ is not yet fully equilibrated,
consistent with the state likely to arise under experimental driving conditions: we label such state here as quasi-adiabatic.
%
We have explicitly verified that driving the system with such $\omega_D$  for half a period (i.e.~up to $t = \pi / \omega_D$), and subsequently removing the drive and waiting sufficient time for the system to relax indeed recovers the corresponding equilibrium state at $g_{\rm min}$.

For such slow quenches, the maximum system density is achieved almost exactly at each local minimum $\Delta g(t)$, while density minima temporally coincide with maxima in $\Delta g(t)$. 
In other words, the density modulation oscillations are in phase with the underlying drive. 
Moreover, the vortex number rapidly decreases acquiring its minimum value at the local $\Delta g$ minimum when the density modulation is also maximised. 


As the oscillation frequency increases beyond the adiabatic limit, 
the overall dynamical evolution becomes faster. To best characterize the underlying dynamics, we consider the dynamical density and vortex number modulation as a function of time scaled to the drive timescale $2 \pi / \omega_D$: this enables evolution graphs to be compared against the same $\Delta g(t)$ graph of Fig.~\ref{fig:quenchprot}(a).
In such scaled time units, the evolution is comparatively slowed down with increased driving frequency. As a result of the faster driving, the system no longer has sufficient time to adjust to the same density maximum within each cycle, with the oscillating density modulation going out of phase (lagging behind) the external periodic driving, by an amount $\phi_{\rm lag}^{n_i}$, corresponding to the phase lag of the $i^{\text{th}}$ trough in $\Delta g(t)$. The local density maximum is thus reached at a time after $\Delta g(t)$ has reached its minimum, when the interaction strength is $g_\text{min}<g(t)<g_c$.

This effect becomes pronounced with faster driving, which leads both to a lower local density maximum being reached, and a longer time delay between driving and density growth. For intermediate external driving frequencies, the first density maximum reached 
is lower than the subsequent ones, with the system evidently requiring multiple driving cycles for the temporally-local density maximum to plateau to its overall maximum attainable value under periodic driving. Moreover, the rapid evolution implies that the overall observed density modulation amplitude per cycle is decreasing with increasing $\omega_D$, so that -- as the density maximum is gradually increasing -- the corresponding density modulation minimum is no longer constrained to coincide with the initial density value: such features are most evident when comparing $\omega_D/2\pi=1/2$Hz (blue) and $\omega_D/2\pi=5/3$Hz (brown) curves in Fig.~\ref{fig:quenchprot}(b).
%
This effect becomes more pronounced with higher values of $\omega_D$, until the driving becomes so rapid that the system has practically no time to adjust.


Such dynamical behaviour can be better understood by looking at the 
corresponding evolution of mean vortex number $\langle N_{v} \rangle$ in the system.
%
As the driving frequency increases, the rate of decrease of vortices in a given driving cycle becomes significantly lower, with the local vortex minimum being significantly delayed from the corresponding $\Delta g$ minimum. The delay of the first local vortex minimum $\phi_\text{lag}^{V_1}$ as measured from the $\Delta g$ minimum at $t=\pi/\omega_D$ is in fact significantly pronounced, and occurs clearly after the corresponding density modulation maximum (which was itself found to phase-lag behind the external driving). 
This can be understood in terms of the additional phase-ordering time required for the annihilation of a vortex-antivortex pair, which is a competing effect to the external drive. 
%
For faster quenches the gradual vortex number decrease occurs over multiple quench cycles and does not necessarily reach $\langle N_{v} \rangle \sim 0$ as evident from the brown quench in Fig.~\ref{fig:quenchprot}(c). 
(In the idealized limiting case of extremely rapid quenches (pink), the system has insufficient time to react, exhibiting only a marginal periodic decrease in vortex number).

The above discussion has identified an evident difference in the evolutions of density and vortex number during the driven quench, identifying two rather different behaviours with respect to the driving frequency.
This points to the existence of a critical (resonant) frequency which separates these two regimes.
We can identify a characteristic relaxation, or `resonant' frequency through the identification of a characteristic timescale in the system.
A relevant timescale in the system is~\cite{McDonald2015}
\begin{align}
    \tau_0=\frac{\hbar}{\gamma|\mu|}\approx 0.4\,\text{s}\,
    \label{eqn:relax1}
\end{align}
This timescale corresponds to a relaxation time towards the symmetric equilibrium above the critical point, while below the critical point it marks a timescale for dynamical instability, through exponential growth of small fluctuations.

\begin{figure}
    \centering
\includegraphics[width=1\columnwidth]{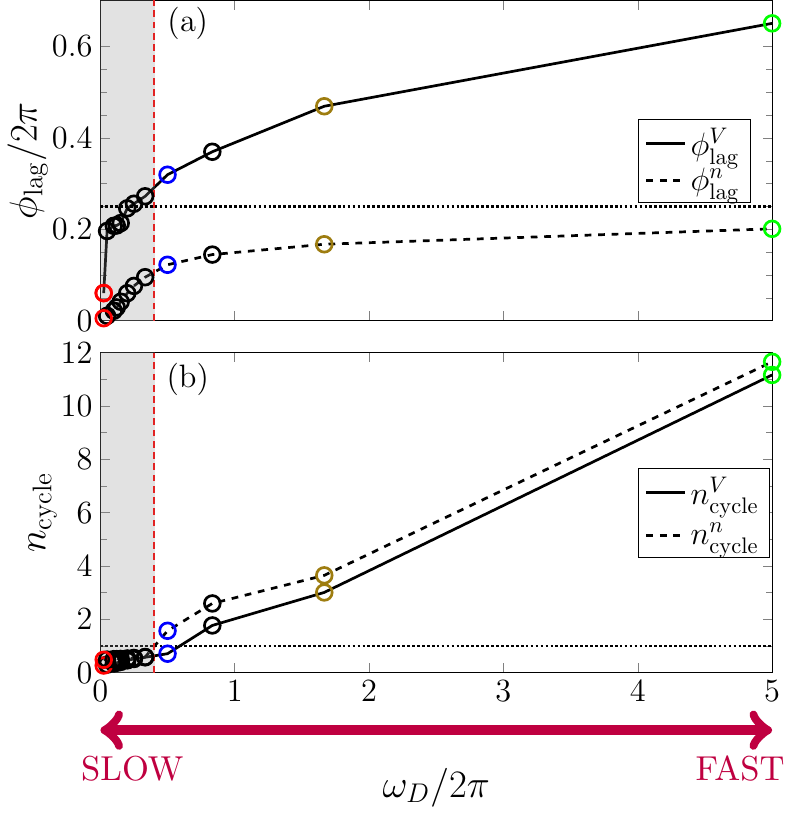}
    \caption{Quench dependent time shifts probed here in the physically interesting range $0.05 \lesssim (\omega_D / \omega_0) \lesssim 10$. (a) The mean phase shift for the density ($\phi^n$) and vortex number ($\phi^V$) between the $i^{\textrm{th}}$ peaks of $\Delta g$ and the peak average density, as a function of $\omega_D$. Error bars are smaller than the marker size. Points corresponding to the quench frequencies considered previously are coloured according to the legend of Fig.~\ref{fig:quenchprot}. (b) Average number of oscillation cycles, $n_\text{cycle}$, between the first quench and density (vortex number) reaching 90\% of its maximum value (10\% of its initial value). The relaxation frequency $\omega_0$ (Eq.~\eqref{eqn:relax}) is marked by a vertical dashed red line, with the grey-shaded area corresponding to the `quasi-adiabatic' regime $\omega_D < \omega_0$.
    Horizontal black lines in (a) and (b) respectively denote the values of $\phi_{\rm lag} / 2 \pi = 1/4$, marking the transition from the observed phase lag occurring within a quarter of the driving cycle (i.e. while the system is within the `coherent' phase), and $n_{\rm cycle} =1$, marking the transition between vortex number counts that are independent and dependent on cycle number.
    For $\omega_D / \omega_0 \gtrsim 100$ (not shown here) the extremely modest response of the system to the drive leads to practically no change in density, or phase.
    }
    \label{fig:taus}
\end{figure}

This timescale in turn implies a characteristic system frequency $\omega_0$, defined as
\begin{align}
    \omega_0=\frac{1}{\tau_0}\approx2\pi\times0.4\,\text{Hz}\,.
    \label{eqn:relax}
\end{align}
This frequency is marked across Fig.~\ref{fig:taus} (and relevant subsequent figures) as a vertical red dashed line, and clearly marks the boundary between quasi-adiabatic driving ($\omega_D < \omega_0$) and dynamically-driven systems ($\omega_D > \omega_0$). The role of this intrinsic `resonant' frequency on momentum spectrum and coherence is further highlighted below.

To characterize the above behaviour quantitatively, we next consider the mean phase lag of both density and vortex evolution as a function of the driving frequency $\omega_D$, as shown in Fig.~\ref{fig:taus}(a). The average vortex number phase lag is greater than the density phase lag for all driving frequencies considered, in keeping with observations from Fig.~\ref{fig:quenchprot}. A phase lag of $\phi_\text{lag}/2\pi=0.25$ corresponds to the external driving parameter crossing of the BKT threshold, which the density phase lag never exceeds, implying the maximum density per cycle is achieved during the coherent phase. However, the minimum in $\langle N_{v} \rangle$ occurs later whilst in the thermal phase, suggesting that the vortex number is not in equilibrium before the next cycle occurs.

Given that the vortex number decay exhibits oscillatory behaviour, and that for a large range of probed driving frequencies it actually requires multiple such oscillations before the vortex number decreases to its steady-state (potentially non-zero) values, it is also interesting to characterize the time taken for the vortex to decay to a near-zero final vortex number. Similarly, the average density also requires several cycles to saturate to its maximum value. 
To quantify these effects, we have here chosen arbitrarily to define a timescale, $n_\text{cycle}$, as the number of quench cycles at which the measured quantity reaches 90\% saturation of its steady-state values. This corresponds to the time the vortex number first reaches 10\% of its initial value, or the density reaches 90\% of its final maximum value. 

\begin{figure*}[!ht]
\centering
\includegraphics[width=0.95\textwidth]{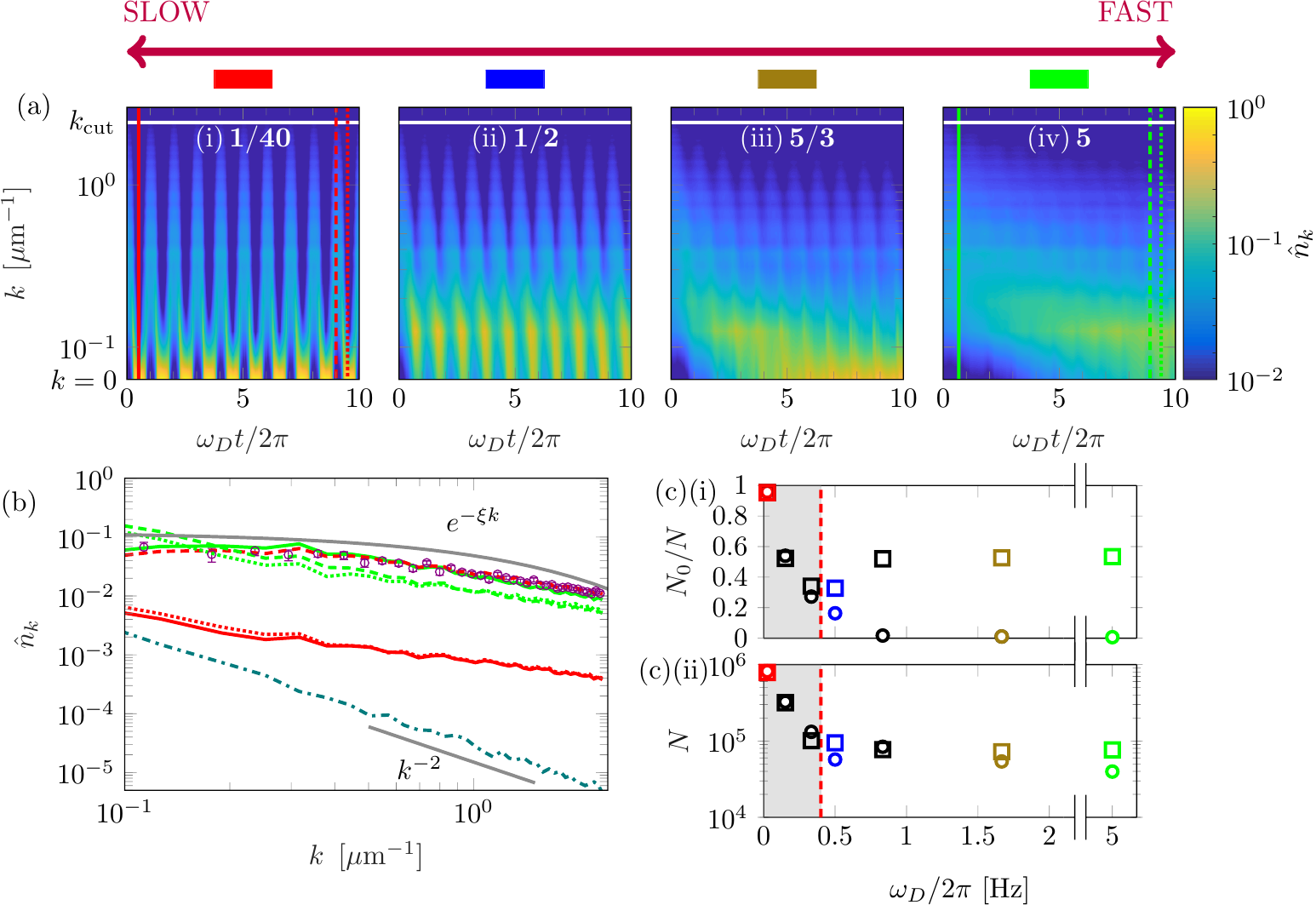}
        \caption{Dynamics of the momentum distribution during the periodic quench. (a) Evolution of the average proportion of atoms in mode $k$, $\hat{n}_k(t)$, defined in Eq.~\eqref{eqn:nknormal}. The value of $\omega_D/2\pi$ (in Hz) is stated at the top of each panel and a coloured block appears above to highlight its relation to previous figures. The solid, dashed and dotted lines in (i) and (iv) pertain to the first maximum, the tenth minimum and the tenth maximum of $\Delta g$, respectively. The occupation of the condensate mode is shown along the bottom parts of these subplots. (b) Slices of momentum density occupation taken through the top panels, showing the profile of $\hat{n}_k$ at different times, given as vertical lines in (a) (i) and (iv). Also shown is the initial $t=0$ distribution prior to the initiation of our periodic quenching (hollow purple circles), with error bars indicative of the magnitude of the error in all of the data shown, and the distribution at equilibrium after a half-quench cycle, followed by a constant $g=g_\mathrm{min}$ post-quench relaxation period (dot-dashed blue line). 
        The expected behaviour in the two limiting cases is shown by solid grey lines: exponential behaviour for a thermal cloud $\sim e^{-\xi k}$ [with a fitted coherence length $\xi=0.91\mu$m $\ll L_x,\,L_y$] (see also Sec.~\ref{sec:g1}~\cite{footnote_g1}) and a $k^{-2}$ scaling for a superfluid system.
        (c)(i) Occupation of the $\hat{n}_{k=0}\equiv N_0/N$ mode for the first density peak (circles) and a late density peak corresponding to the time $\gamma t_*=0.12$s (squares) [see later]. (c)(ii) Total atom number evaluated at the same times as (c)(i). A vertical red line depicting the relaxation frequency $\omega_0$ (Eq.~\eqref{eqn:relax}) appears in both panels of (c).}
    \label{fig:momentumspace}
\end{figure*}

Apart from the very slow quenches, Fig.~\ref{fig:taus}(b) shows a near-linear increase in the number of cycles required for the system to reach the steady-state as a function of driving frequency. Moreover, even though it may take multiple cycles to reach that value, the linear-like curve implies a near constant duration of time to saturation ($2\pi n_\text{cycle}/\omega_D$) for $\omega_D \gtrsim \omega_0 $. This figure thus confirms the two distinct regimes either side of $\omega_D \sim \omega_0$ probed here in the interesting range $0.05 \lesssim (\omega_D / \omega_0) \lesssim 10$. For relatively slow driving $\omega_D < \omega_0$, the vortex number and average density saturate within a single quench cycle, suggesting maximum coherence has been reached within one cycle and the vortex number and density evolution are only dependent on the time during a cycle when they are measured, and not the total number of cycles completed. However, when driving faster than $\omega_0$, the system saturates after multiple quench cycles,
with extreme limits of $(\omega_D / \omega_0) \gg 10$ (not shown here) leading to only very small variations from the initial (incoherent) state.


We now turn our attention to the evolution of the momentum spectrum as a function of different driving frequencies.

\subsection{Momentum Evolution and Condensate Mode Dynamics}

The density analysis considered above focussed on the entire classical field region, which encompasses both the condensate--which requires macroscopic occupation of the lowest lying momentum mode--and those higher-lying macroscopically occupied modes affected by its presence. To quantify the emerging (quasi)condensate dynamics, we thus consider the evolution of the momentum distribution. 

Firstly, we define the fractional momentum occupation of mode $k$ via
\begin{align}
    \hat{n}_k(t) = \frac{ \langle \tilde{n}_k(t) \rangle_{\mathcal{N}} }{ \sum\limits_{k} \langle \tilde{n}_k(t) \rangle_{\mathcal{N}} }\,,
    \label{eqn:nknormal}
\end{align}
where $\langle\cdot\rangle_\mathcal{N}$ denotes averaging over stochastic realisations.
The evolution of the normalized momentum density $\hat{n}_k$ is plotted in Fig.~\ref{fig:momentumspace}(a)  for different driving frequencies from slow (left, 
$\omega_D/\omega_0 \sim 0.05$)
through to fast (right, $\omega_D/\omega_0 \sim 10$) quenches.
For a truly adiabatic evolution, proceeding slowly enough that the system passes through {\em fully equilibrated} states during its entire evolution, one would expect to see the periodic emergence of a `condensate' mode, in the sense of a highly populated $k = 0$ mode
Indeed, this emerges for
the slowest quasi-adiabatic periodic quench [panel (a)(i)], whose modes up to $k \sim 0.1 \mu$m are maximally populated at times $t = (2\pi/\omega_D )(2m+1)/2$ (where $m$ is an integer), when $\Delta g$ has a local minimum.

This regime clearly shows a periodic reversible condensate formation process, qualitatively similar to the behaviour observed in harmonic dimple microtraps cycled across the phase transition~\cite{stamperkurn1998reversible,stoof2001dynamics}.
As the quench frequency increases beyond $\omega_0 \sim 2 \pi \times 0.4$ Hz we expect coherence to form over multiple quench cycles. Indeed for $\omega_D \sim 1.25 \omega_0$ [panel (a)(ii)], we find the initial momentum peak after a single quench cycle to be significantly broadened around $k \sim 0$, consistent with the existence of multiple highly-occupied modes (and the gradual emergence of a quasi-condensate).
Further increasing the driving frequency leads to coherence only appearing after multiple cycles, consistent with earlier findings, due to the more gradual decrease of vortex numbers which--for faster quenches--becomes significant only after multiple driving cycles through the critical point.


Example slices of $\hat{n}_k$ vs.~$k$ (for $k>0$), at extreme limits of $\Delta g$ are shown in Fig.~\ref{fig:momentumspace}(b) for both a relatively slow (red; corresponding to a(i)) and a relatively fast (green; corresponding to a(iv)) drive. As a guide for the eye, solid grey lines show the following limiting behaviour: a thermal field is expected to follow an $e^{-\xi k}$ scaling, where $\xi$ is the coherence length from the first order correlation function $g^{(1)}(r)$ for a thermal cloud~\cite{footnote_g1}, and a fully equilibrated 2D Bose gas is expected to exhibit a $k^{-2}$ scaling~\cite{pitaevskii2003bose}. Our findings show that the driven states oscillate between these limiting cases. As a comparison, we show the extracted momentum distribution for the thermal field at $t=0$ (purple circles), and of the Bose gas in equilibrium showing the expected $k^{-2}$ behaviour (blue dash dotted), and find excellent agreement to the expected scaling.

Considering first the slow driving case [case (a)(i)], we note that the system exhibits a monotonically-decaying $\hat{n}_k$ [Fig.~\ref{fig:momentumspace}(b)] with more than 95\% condensate fraction [Fig.~\ref{fig:momentumspace}(c)] both at the first (solid red line) and subsequent $\Delta g$ minima (dotted red line). This is in contrast to the corresponding behaviour at $\Delta g$ maxima (dashed red line) which shows decreasing occupation as $k \rightarrow 0$. 


Considering now a much faster quench [case a(iv)], the first $\Delta g$ minimum [solid green line] is clearly not condensed, as evident from the low $k=0$ occupation in Fig.~\ref{fig:momentumspace}(c)(i) [green open circle], and still resembles the initial thermal field distribution at $t=0$. After many driving cycles, such a rapidly-driven system develops clear evidence of non-equilibrium condensation, with increasing occupation into the lower $k$ modes [dotted and dashed green lines] and $\sim50\%$ occupation into the $k=0$ mode (Fig.~\ref{fig:momentumspace}(c)(i)), distinct from the expected thermal distribution and with little variation between the maximum (dotted) and minimum (dashed) cases of $\Delta g$. 
Remarkably, for fast quenches the system maintains a coherent fraction regardless of the drive. 

In all cases, it's worth noting that the spectrum at large $k$ acquires a power-law behaviour with an exponent resembling  $\sim-1$~\cite{footnote_scaling}, clearly distinct from both expected thermal and Bose gas scalings. 
We attribute such numerically observed `anomalous' scaling to the highly non-equilibrium nature of the driven system, a feature that could be tested in experiments.

The maximum of the condensate fraction as a function of driving frequency is shown in Fig.~\ref{fig:momentumspace}(c)(i), and the total atom number in (c)(ii). The open circles show the condensate fraction measured at the first peak in the total density, whereas the open squares show the condensate fraction measured at an arbitrary peak at sufficiently late times. For faster driving frequencies the condensate fraction {\em increases} over time, as a result of the repeated crossing of the phase transition,
thus leading to a non-equilibrium steady-state with {\em irreversible} growth.
This can be attributed to the total time spent above the phase transition, $\pi/\omega_D$, being smaller than the relaxation time $\tau_0$, and hence a small fraction of the atoms remain Bose condensed after each repeated crossing of the phase transition. This effect disappears both for small driving frequencies, where the condensate fraction follows the equilibrium value expected from an instantaneous evaluation of the dissipative GPE, and for very rapid frequencies
($\omega_D / \omega_0 \gtrsim 100$ -- beyond the scale of this graph)
which do not allow for any noticeable coherence to form even after multiple quench cycles. In this limit the condensate feels a time averaged drive, $\langle g(t)\rangle_t=g_c$, and the c-field remains incoherent. 

Interestingly, the smallest condensate fraction occurs around $\omega_D \approx \omega_0$, when the competing driving and relaxation are in {\em resonance} with one another, 
leading to a highly non-equilibrium partly-superfluid state,
an effect we explore in more detail in Sec.~\ref{sec:resonance}.

\subsection{Evolution of Phase Coherence}\label{sec:g1}

We now discuss the effect of driving on the spatial coherence of the system. 

\begin{figure}
    \centering
\includegraphics[width=1\columnwidth]{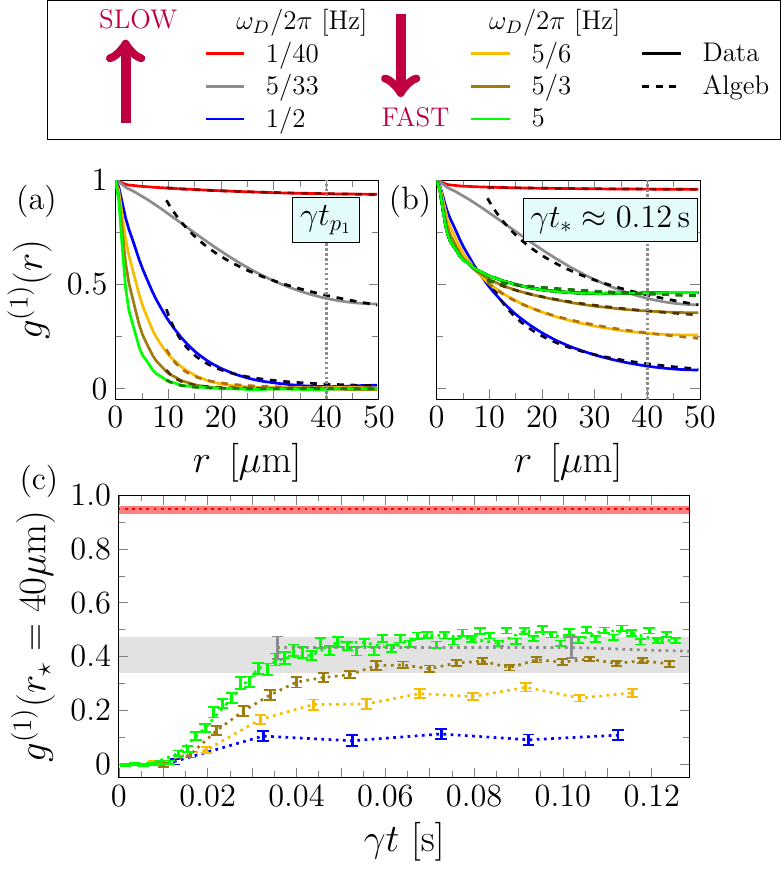}
\caption{Growth of coherence through repeated quenches. (a) First order correlation function, $g^{(1)}(r)$, evaluated at $t=t_{p_1}$, the time of the first peak in the average density in Fig.~\ref{fig:quenchprot}. Each curve is fitted with an algebraic function, $g^{(1)}(r) \sim r^{-\alpha}$, plotted using a dashed line. (b) Correlation function evaluated at the density peak closest to $\gamma t_* = 0.12$s. (c) Correlation function evaluated at $r_\star=40 \mu$m, 
[indicated by the dashed vertical grey lines in panels (a)-(b)],
as a function of time. Error bars are located where $g^{(1)}$ is  measured at a peak in the average density. Data for $\omega_D/2\pi=1/40$ (red) and $5/33$ (grey) Hz extend far beyond $\gamma t=0.12$s, with corresponding red/grey bands denoting error bars over the total range covered. Plotted error bars indicate one standard deviation of $g^{(1)}(r)$.
}
    \label{fig:g1}
\end{figure}

For a homogeneous 2D system, the spatial correlation function, or normalised first order correlation function, defined by \cite{pitaevskii2003bose}:
\begin{align}
    g^{(1)}(r)=\frac{\langle\Phi^*(r_0)\Phi(r_0+r)\rangle}{\sqrt{\langle|\Phi(r_0)|^2\rangle\langle|\Phi(r_0+r)|^2\rangle}}\,,
    \label{eqn:g1}
\end{align}
is known (at equilibrium) to exhibit a transition from exponential (above the critical region) to algebraic-order decay (below the critical region), a characteristic of the BKT phase transition~\cite{Nazarenko2014,prokofev_two-dimensional_2002,foster2010vortex,gawryluk_19,comaron_19,hadzibabic2006berezinskii}.
We have indeed verified such behaviour in the steady-state profiles for the two limiting equilibrium cases $g=g_\text{max}$ and $g=g_\text{min}$. For a thermal cloud $g^{(1)}(r)\sim e^{-r/\xi}$, where $\xi$ is the coherence length
[see 
also Fig.~\ref{fig:momentumspace}(b)].
Here we are instead interested in investigating the effect of the periodic driving on the emerging maximum spatial phase coherence. 
With this in mind, we perform a detailed analysis of the system coherence when $\Delta g =-0.9$. To account for the fact that the system does not necessarily relax after a single driving cycle (for faster driving), we consider the spatial phase coherence  at two times. These correspond to: (a) the end of the first driving half-period ($t_{p_{1}}$) when the system reaches its minimum $\Delta g$ value after a single phase transition crossing -- this is shown in  Fig.~\ref{fig:g1}(a). This corresponds to the same point in the driving cycle irrespective of $\omega_D$, thus amounting to different physical times.
(b) We also measure $g^{(1)}$ at a fixed absolute late time for all $\omega_D$, corresponding to a sufficiently long time such that all considered cases have already acquired their maximum steady-state values. Specifically: we investigate $g^{(1)}(r)$ at the $\Delta g$ trough at, or just after the time $\gamma t_* \approx 0.12$s:
such a time corresponds to the first trough in the limiting case of quasi-adiabatic driving ($\omega_D = 2 \pi (1/40)$ Hz $=0.05 \omega_0$) and the 60$^\text{th}$ trough, at which a non-equilibrium steady-state has been reached for the ultrafast quench $\omega_D = 2 \pi \times5\text{Hz} = 10 \omega_0$ ]. This is shown in Fig.~\ref{fig:g1}(b).

Fig.~\ref{fig:g1}(a) shows that for the quasi-adiabatic driving [$\omega_D = 2 \pi \times (1/40)$Hz], the system exhibits very strong spatial coherence at the first $\Delta g$ minimum, which is perfectly fit with an algebraic-decay function $g^{(1)}(r) \sim r^{-\alpha}$, where $\alpha = 6.8\times10^{-3}$. This correlation is actually identical in all subsequent peaks with the same driving.
Although this is highly coherent, consistent with the complete absence of vortices in most numerical realizations, we nonetheless note here once again that this system is not yet fully equilibrated.
This is because the emerging state is found to exhibit a momentum spectrum distinct from the expected $k^{-2}$ equilibrium spectrum, whereas a $k^{-2}$ spectrum does emerge upon allowing this same system to relax (without further driving) for a significantly longer timescale at $\Delta g = -0.9$.

Increasing the driving frequency leads to the establishment of less coherence in the system after a single cycle, with coherence 
nonetheless
building up gradually after multiple cycles. This can be seen by comparing the computed $g^{(1)}(r)$ between Figs.~\ref{fig:g1}(a) and (b). 
Although very fast quenches lead to very small coherence after a single drive half-cycle, probing the correlation function instantaneously at $g=g_\text{min} \ll g_{c}$ nonetheless still facilitates a very good algebraic-decay fit (but with a higher value of $\alpha$, implying a highly  non-equilibrium state). Thus, in order to compare with the maximal coherence imparted at steady-state, we analyze all plots here in terms of an algebraic-decay fit.

The increase of the (algebraic) decay rate of $g^{(1)}(r)$ with increasing  $\omega_D$ seen in Fig.~\ref{fig:g1}(a) should come as no surprise, since the build-up of significant coherence requires the decay of all vortices in the system, whereas all probed quenches except the quasi-adiabatic (red) one still have at least some vortices present at the first half-cycle [see Fig.~\ref{fig:quenchprot}(c)]. Remarkably, even the presence of a few vortices is enough to significantly destroy the system's overall phase coherence, as evident from the blue curves [$\omega_D = 2 \pi \times (1/2)$ Hz] in Fig.~\ref{fig:g1}(a) and Fig.~\ref{fig:quenchprot}(c).

An interesting and rather distinctive feature arises when looking at corresponding late-time peaks shown in Fig.~\ref{fig:g1}(b), after allowing the periodically-driven system to reach its non-equilibrium steady-state, following a sufficiently large number of drive cycles, and repeated crossings of the phase transition in both directions. We have already shown earlier that for sufficiently fast driving, the vortex number curve decreases periodically in an oscillating manner, but with a rapidly decaying envelope, 
requiring multiple cycles for a significant decrease of its mean vortex number $\langle N_{v}\rangle$. As a result, coherence builds up gradually in such systems, and this was to be expected. The interesting emerging feature here is that at steady-state, there is an optimum driving frequency, above which the final acquired system coherence (at one of the $\Delta g$ peaks) at steady-state grows over longer distances again.

This is easily seen by comparing Fig.~\ref{fig:g1}(a)-(b). Focussing at the late time $\gamma t_* \sim 0.12$s, we see that increasing the frequency (in Hz) from $\omega_D / 2 \pi = 1/40$ (red) through $5/33$ (brown) to $1/2 \sim O(\omega_0 / 2 \pi)$ (blue), $g^{(1)}(r)$ decays faster spatially with increasing drive frequency. However, driving the system even faster than $\omega_0$ periodically across a sufficiently large number of phase-transition cycles leads to an {\em increase} in the system phase coherence, with the spatial coherence at the maximum distance (half the box length) of the $\omega_D / 2 \pi = 5$ Hz (green) being comparable to that of $\omega_D / 2 \pi = 5/33$ Hz (grey), which is driven 33 times slower (and whose coherence does not change between consecutive peaks, having already saturated at its maximal coherence at the end of the first half-cycle).

Such behaviour becomes more evident in Fig.~\ref{fig:g1}(c), when examining the temporal evolution of the value of the phase correlation function near the edge of the system, at $g^{(1)}(r_\star=40\mu$m). Slow quenches ($\omega_D/2\pi<0.5$ Hz) allow maximal coherence to be effectively established already after a single quench half-cycle: this is indicated by the red/grey horizontal lines/bands in Fig.~\ref{fig:g1}(c). For faster driving than that, the system initially reacts rather slowly, with the coherence growth rate gradually picking up around $\gamma t \sim 0.01$s, eventually saturating at a higher value (well before the $\gamma t \sim 0.12$s used for the comparison in Fig.~\ref{fig:g1}(b)).
Interestingly the extra growth of coherence after multiple cycles is very small around the frequency $\omega_D / 2 \pi \sim 1/2$ Hz
(blue curve), but increases noticeably for quenches few times faster than that, as evident from the enhanced values shown by the yellow, brown and green curves.

\subsection{Resonances with intrinsic timescales}\label{sec:resonance}

\begin{figure}[t]
    \centering
 \includegraphics[width=1\columnwidth]{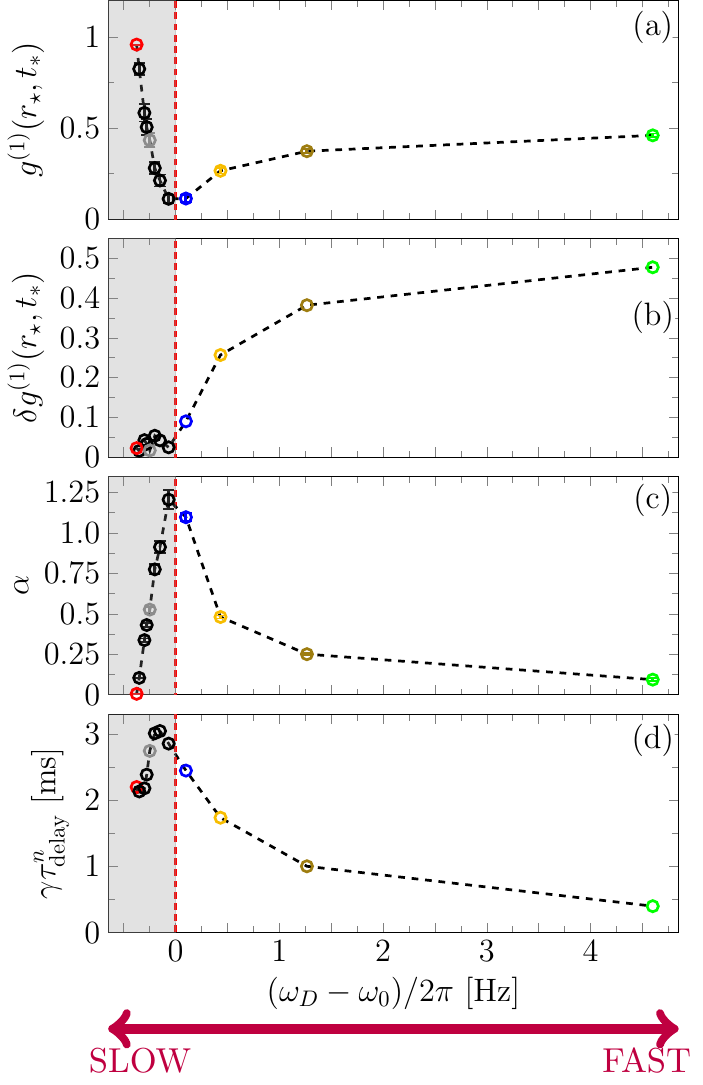}
    \caption{Resonances of observables with the relaxation time. (a) Correlation function evaluated at $r_{\star} = 40\mu$m, $\gamma t_* \approx 0.12$s, see Fig~\ref{fig:g1}. (b) Percentage change in $g^{(1)}(r_\star,t)$ from $t_1$ to $t_*$ (early to late times). (c) Exponent from algebraic fits of Fig.~\ref{fig:g1}(b). (d) Average time delay in the average peak density.}
    \label{fig:analysis}
\end{figure}

The behaviour of the spatial correlation function is further characterized in comparison to the relaxation frequency $\omega_\text{0}=2\pi\times0.4$ Hz in Fig.~\ref{fig:analysis}. Here we plot (a) the steady-state value ($t=t_*$) of $g^{(1)}$ near the box edge ($r=r_\star$), (b) 
the fractional increase of its value at $t_*$, compared to that at its first peak, 
(c) the value of the power-law decay exponent $\alpha$, and (d) the density phase lag $\gamma \tau^{n}_\text{delay}= 2\pi\phi^n_\text{lag}/\omega_D$ (a) (corresponding to Fig.~\ref{fig:taus}(a), but now plotted in terms of average delay time which offers a different perspective).
This figure clearly highlights the importance of the critical driving frequency, corresponding to `resonant' driving distinguishing the slow and fast driving regimes.
This resonant driving frequency is well explained in terms of the frequency $\omega_{0}$. Looking at the various subplots, we easily infer that maximal coherence is achieved in the quasi-adiabatic regime (leftmost, red points in all subplots), with $g^{(1)}(r_\star=40\mu m,\, \gamma t_*=0.12s)$ $\sim 0.96$ and $\alpha = 6.8\times 10^{-3}$, which is, as expected for such a low-temperature state with $T/T_{BKT} \sim 0.15$, much less than the value $\alpha = 0.25$ occurring at equilibrium for $g=g_c$ \cite{Nelson1977}.

As the system is driven faster, the late-time coherence at the system edge rapidly decreases (down to $\sim 0.11$) [Fig.~\ref{fig:analysis}(a)], with such 
value being already reached after a single quench half-cycle:
the latter can be inferred by looking at the fractional change in the value of the correlation function at $r_\star$ from the first peak up to a late converged peak at $t_*$.
Correspondingly, faster driving leads to an increasing delay (phase lag) of density growth, as shown in Fig.~\ref{fig:analysis}(d).
Interestingly, an algebraic power-law fit $g^{(1)}(r,\,t) \sim r^{-\alpha}$ of the correlation function at  $\gamma t_* \sim 0.12$s -- when the system clearly possesses a non-zero occupation of the $k=0$ mode -- leads to a significant increase in the value of $\alpha$, much exceeding the equilibrium value of (1/4). 
This latter observation provides
%
strong evidence of the non-equilibrium nature of the achieved steady-state, 
and points to the strong interplay between the system's external driving across the phase transition (quantified in $\omega_D$), and its ability to adjust (quantified in $\omega_0$) to an averaged behaviour across two vastly different equilibrium states located well above and well below the critical region.

When driving the system even faster than that, the system loses its ability to adjust rapidly, as it starts experiencing an averaged quench effect. The overall steady-state coherence at the trap edge starts increasing again with increasing frequency [Fig.~\ref{fig:analysis}(a)], with an increase of as much as 50\% over its initial (but rather low) acquired value at the first quench half-cycle [Fig.~\ref{fig:analysis}(b)]. As the system has less and less time to react to the external driving, it effectively only exhibits a time-averaged effect, and so becomes less non-equilibrium, consistent with a decreasing value of $\alpha$ with faster driving $\omega_D > \omega_0$ [Fig.~\ref{fig:analysis}(c)]. As a result of the rapid driving, the overall density evolution time-delay now also decreases significantly (in `absolute' time $\gamma t$ [Fig.~\ref{fig:analysis}(d)]), which is testament to the very tiny increase of the system density facilitated.

Nonetheless, it is important to note that this delay time is actually a significant fraction of the drive time, as evident from Fig.~\ref{fig:taus}(a), revealing that the density delay time for the fastest quench considered here is 20\% of the quench period, while the vortex delay time [not shown here, see Fig.~\ref{fig:taus}(a)] exceeds 60\% of the period.

Much faster driving than that ($\omega_D / \omega_0 \gtrsim 100$) [not shown here] leads to a practically monotonic time-averaged evolution and a minor decrease of about 30\% in vortex number, with an associated increase of less than 10\%in the population of the $k=0$ mode, indicating a vortex-filled, low-coherence, non-equilibrium state.
Moreover, driving at an idealized `infinite' frequency $\omega_D / \omega_0 \sim 1000$ was found to lead to practically no change to the system's initial configuration.

\section{Conclusions}\label{sec:conclusion}

We have considered the dynamical response of a homogeneous two-dimensional ultracold Bose gas under periodic quenches of its interaction strength through the Berezinskii-Kosterlitz-Thouless phase transition at a driving frequency $\omega_D$. 
We have identified an intrinsic system response frequency $\omega_0$ and demonstrated that resonant driving leads to a highly non-equilibrium state exhibiting only limited coherence growth, compared to when driving the system on either side of this resonance.
Focusing on the most interesting regime of driving at a frequency within one order of magnitude from the resonant value, we characterized the system response in terms of the driving frequency, by analysing the evolution of densities, vortices, condensate fractions, spectra and coherence.

Specifically, we identified two distinct driving regimes of experimental relevance: 
Driving at a frequency $\omega_D$ much smaller than $\omega_0$ gives rise to quasi-adiabatic dynamics, with maximum coherence achieved after a single quench half-cycle, even if the system has not had sufficient time to fully equilibrate in momentum space (with the limiting case of extremely slow drive, $\omega_D \rightarrow 0$, corresponding to the adiabatic regime).  Under (quasi)-adiabatic conditions it is possible to analytically describe the evolution of the average density.

In the opposite regime, driving the system faster than the system response frequency leads to a dynamically-driven non-equilibrium system, whose coherence grows gradually through multiple quench cycles. Interestingly, such driving can lead to a significant condensate fraction and enhancement in coherence: for example, even for large driving frequencies $\omega_D/\omega_0 \sim 10$, the coherence can grow up to 50\% over the range of the box, despite only achieving $<1\%$ coherence after the first quench.
As expected, quenching much faster than that (e.g.~$\omega_D / \omega_0 \gtrsim 100$) leads to practically no generation of coherence, with the system entering an incoherent, vortex-filled steady-state.

Our study extends earlier works on cyclic phase transition crossing, and may help guide future non-equilibrium driven quench experiments and lead to the generation of interesting non-equilibrium steady-states of ultracold atoms with partial superfluidity.





Data supporting this publication is openly available under a
Creative Commons CC-BY-4.0 License on the data.ncl.ac.uk site \cite{data}.

\

\section*{Acknowledgements}

We thank Gerasimos Rigopoulos for discussions. KB acknowledges support from a Newcastle University summer studentship. TB thanks EPSRC Doctoral Prize Fellowship Grant No.~EP/R51309X/1. We acknowledge financial support from the Quantera ERA-NET cofund project NAQUAS through the Engineering and Physical Science Research Council, Grant No.~EP/R043434/1. 

\bibliographystyle{apsrev}
\bibliography{biblio}

\begin{thebibliography}{112}
\bibitem[{foo({\natexlab{a}})}]{footnote1}
\bibinfo{note}{Parameters used here were inspired by potentially accessible
  regimes in future experiments at Cambridge (Julian Schmitt, Private
  Communication)}.
\bibitem[{foo({\natexlab{b}})}]{footnote2}
\bibinfo{note}{Reducing $g$ at fixed $T$ and $\mu$ in the SPGPE causes the
  critical temperature to increase -- distinct to what would happen if the
  quench were done at constant density. Specifically here:
  $T_\text{BKT}(g=1.9g_c) = 13.8$nK, while $T_\text{BKT}(g=0.1g_c) = 371$nK
  (with the corresponding c-field atom numbers at equilibrium being $3 \times
  10^{4}$ and $8.2 \times 10^{5}$). 
  corresponding values $T\ll T_\text{BKT}(\tilde{g}=0.00183)$ and $T\gg
  T_\text{BKT}(\tilde{g}=0.03483)$]}.
\end{thebibliography}

\end{document}